\documentclass[journal,twoside,web]{ieeecolor}
\usepackage{tmi}
\usepackage{cite}
\usepackage{amsmath,amssymb,amsfonts}
\usepackage{algorithmic}
\usepackage{graphicx}
\usepackage{textcomp}
\usepackage{hyperref}
\usepackage{multirow}
\usepackage{float}
\usepackage{booktabs}
\hypersetup{
    colorlinks=true,
    linkcolor=black,
    citecolor=black,
    urlcolor=black
}

\def\BibTeX{{\rm B\kern-.05em{\sc i\kern-.025em b}\kern-.08em
    T\kern-.1667em\lower.7ex\hbox{E}\kern-.125emX}}
{R.WU \MakeLowercase{\textit{et al.}}: Spatial-Angular Representation Learning for Diffusion MRI}
\begin{document}
\title{Spatial-Angular Representation Learning for High-Fidelity Continuous Super-Resolution in Diffusion MRI}
\author{Ruoyou Wu, \IEEEmembership{Student Member, IEEE}, Jian Cheng, Cheng Li, \IEEEmembership{Member, IEEE}, Juan Zou, Wenxin Fan, Hua Guo, Yong Liang, \IEEEmembership{Member, IEEE}, and Shanshan Wang, \IEEEmembership{Senior Member, IEEE}
\thanks{Manuscript received **; accepted **. This work was supported in part by the National Natural Science Foundation of China under Grant 62222118 and Grant U22A2040, in part by Shenzhen Science and Technology Program under Grant RCYX20210706092104034 and Grant JCYJ20220531100213029, in part by Shenzhen Medical Research Fund under Grant B2402047, in part by Key Laboratory for Magnetic Resonance and Multimodality Imaging of Guangdong Province under Grant 2023B1212060052, and in part by the Youth Innovation Promotion Association CAS. (\textit{Corresponding author: Shanshan Wang.})}
\thanks{Ruoyou Wu is with the Paul C. Lauterbur Research Center for Biomedical Imaging, Shenzhen Institutes of Advanced Technology, Chinese Academy of Sciences, Shenzhen 518055, China, and with the Pengcheng Laboratory, Shenzhen 518055, China, and also with the University of Chinese Academy of Sciences, Beijing 100049, China (e-mail: ry.wu@siat.ac.cn). }
\thanks{Jian Cheng is with the School of Computer Science and Engineering, Beihang University, Beijing 100191, China (e-mail: jian\_cheng@buaa.edu.cn).}
\thanks{Cheng Li, Wenxin Fan, and Shanshan Wang are with the Paul C. Lauterbur Research Center for Biomedical Imaging, Shenzhen Institutes of Advanced Technology, Chinese Academy of Sciences, Shenzhen 518055, China (e-mail: cheng.li6@siat.ac.cn; wx.fan@siat.ac.cn; sophiasswang@hotmail.com).}
\thanks{Juan Zou is with the School of Physics and Electronic Science, Changsha University of Science and Technology, Changsha 410114, China (e-mail: zjuan@csust.edu.cn).}\thanks{Hua Guo is with the Center for Biomedical Imaging Research, Department of Biomedical Engineering, School of Medicine, Tsinghua University, Beijing 100084, China (e-mail: huaguo@tsinghua.edu.cn)}
\thanks{Yong Liang is with the Pengcheng Laboratory, Shenzhen 518055, China (e-mail: liangy02@pcl.ac.cn).}}

\maketitle

\begin{abstract}
Diffusion magnetic resonance imaging (dMRI) often suffers from low spatial and angular resolution due to inherent limitations in imaging hardware and system noise, adversely affecting the accurate estimation of microstructural parameters with fine anatomical details. Deep learning-based super-resolution techniques have shown promise in enhancing dMRI resolution without increasing acquisition time. However, most existing methods are confined to either spatial or angular super-resolution, limiting their effectiveness in capturing detailed microstructural features. Furthermore, traditional pixel-wise loss functions struggle to recover intricate image details essential for high-resolution reconstruction. To address these challenges, we propose SARL-dMRI, a novel Spatial-Angular Representation Learning framework for high-fidelity, continuous super-resolution in dMRI. SARL-dMRI explores implicit neural representations and spherical harmonics to model continuous spatial and angular representations, simultaneously enhancing both spatial and angular resolution while improving microstructural parameter estimation accuracy. To further preserve image fidelity, a data-fidelity module and wavelet-based frequency loss are introduced, ensuring the super-resolved images remain consistent with the original input and retain fine details. Extensive experiments demonstrate that, compared to five other state-of-the-art methods, our method significantly enhances dMRI data resolution, improves the accuracy of microstructural parameter estimation, and provides better generalization capabilities. It maintains stable performance even under a 45$\times$ downsampling factor.
\end{abstract}

\begin{IEEEkeywords}
Diffusion MRI, continuous super-resolution, implicit neural representations, spherical harmonics, high-fidelity.
\end{IEEEkeywords}

% section 1 Introduction
\section{Introduction}
\label{sec:introduction}
\IEEEPARstart{D}{iffusion} magnetic resonance imaging (dMRI) is a widely used non-invasive medical imaging technique that indirectly studies and characterizes the microstructure and functional connectivity of brain tissue by evaluating the diffusion of water molecules in human tissues \cite{descoteaux2007regularized, tournier2011diffusion, martin2025conditional,spagnolo2024down}. It is also widely used in the study of neurological disorders affecting the brain, such as epilepsy, brain malformations, and autism\cite{cascio2007diffusion,sun2024disturbed,eriksson2001diffusion,shen2024exploring}. For example, studies have shown that during brain development, the fractional anisotropy (FA) in classical diffusion tensor imaging (DTI) \cite{le2001diffusion} tends to increase, while mean diffusivity (MD) shows a decreasing trend\cite{dubois2014early,cascio2007diffusion}. 

Although dMRI has seen widespread application, its resolution is restricted by the scanning constraints of imaging systems and inherent system noise, characterized by low spatial and angular resolution. These limitations compromise the accuracy of microstructural parameter estimation, such as FA and MD, in fine anatomical structures\cite{yin2016unified}. Specifically, acquiring dMRI data typically requires a trade-off between resolution, signal-to-noise ratio (SNR), and acquisition time\cite{chen2019xq}. When the acquisition time is fixed, the spatial resolution of dMRI data is limited by low SNR. In dMRI, accurate estimation of microstructural parameters typically requires the acquisition of diffusion-weighted (DW) images across multiple b-values and diffusion directions, which prolongs scan times and increases the risk of patient motion\cite{yin2018joint,luo2022diffusion}. Therefore, acquiring high-resolution dMRI data without increasing acquisition time continues to be a challenge in clinical practice\cite{scherrer2011super}.

Deep learning-based image super-resolution methods offer an effective way to enhance the spatial resolution of dMRI data without significantly reducing SNR or increasing acquisition time\cite{wang2024knowledge,du2020super, alexander2017image,tanno2021uncertainty,chatterjee2021shuffleunet,wu2022arbitrary,zhu2023super,guo2023semi, zhao2024super,wu2024csr}. These methods are post-processing approaches, making them independent of the acquisition protocol\cite{yin2018joint}. Some studies directly leverage neural network models from computer vision to improve the spatial resolution of dMRI data\cite{luo2022diffusion,chatterjee2021shuffleunet,guo2023semi,albay2018diffusion}. For instance, Albay et al. \cite{albay2018diffusion} utilize a generative adversarial network (GAN) model to enhance the spatial resolution of dMRI data. Rather than merely interpolating existing data, this method learns a data-driven generative mapping. Furthermore, sub-pixel convolution \cite{shi2016real} has been employed to improve the spatial resolution of dMRI data. Luo et al. \cite{luo2022diffusion} propose a GAN with sub-pixel convolution for dMRI super-resolution, which enhances the quality of generated images through its integration into the generator. Chatterjee et al. \cite{chatterjee2021shuffleunet} effectively improve image resolution by incorporating sub-pixel convolution into their model, thereby avoiding checkerboard artifacts. 
In addition, graph convolutional neural networks\cite{te2018rgcnn} have been used to capture information from the diffusion domain to help improve the spatial resolution of dMRI data\cite{zhu2023super, zhao2024super, ma2021hybrid}. For example, Ma et al. \cite{ma2021hybrid} propose a hybrid graph convolutional neural network that improves the spatial resolution of dMRI data by incorporating a graph model into a deep network. Uncertainty quantification has also been incorporated into dMRI super-resolution to enhance image reliability and improve model robustness \cite{tanno2021uncertainty,wang2023uncertainty}. For instance, Tanno et al.\cite{tanno2021uncertainty} utilize a heteroscedastic noise model to explain intrinsic uncertainty and approximate Bayesian inference to address parameter uncertainty, combining these approaches to quantify the predictive uncertainty of the output images. 

While these methods effectively enhance the spatial resolution of dMRI data using deep learning without increasing acquisition time, they rarely address the angular resolution of dMRI data. Performing super-resolution only in the spatial domain is insufficient to improve diffusion direction information within each voxel, resulting in inaccurate microstructural parameter estimation\cite{yin2018joint}. Additionally, neglecting angular resolution overlooks the relationships between diffusion images in different directions\cite{yin2016unified}. Some researchers have attempted to use deep learning to obtain high angular resolution diffusion images from low angular resolution dMRI data. These methods can generally be categorized into two types: super-resolution of DW images \cite{yin2019fast,lyon2022angular,chen2023super,lyon2024spatio,weiss2021towards,suzuki2024high} and task-driven super-resolution tailored for downstream applications\cite{liu2023accelerated,wang2024ultrafast,karimi2022diffusion,tian2020deepdti,li2021superdti,qin2021super}. Relatively speaking, the former is suitable for various microstructural parameter estimation tasks, but its accuracy may be constrained. The latter is often constrained by specific downstream tasks, thereby limiting its applicability \cite{lyon2024spatio}. For the super-resolution of DW images, Yin et al. \cite{yin2019fast} use a one-dimensional convolutional autoencoder to learn the relationships between different diffusion directions. Still, they neglect the spatial correlations in dMRI data. Considering the unique geometric structure of dMRI data, Lyon et al. \cite{lyon2024spatio} utilize a parameterized continuous convolutional framework for angular super-resolution. Additionally, they introduce Fourier feature mapping, global coordinates, and domain-specific contextual information to improve the model’s performance. For task-driven super-resolution, Li et al. \cite{li2021superdti} develop a deep learning-based reconstruction framework for robust diffusion tensor imaging and fiber tractography. Qin et al. \cite{qin2021super} extend q-space deep learning (q-DL) to tissue microstructure estimation, proposing the super-resolved q-DL (SR-q-DL) method. Additionally, they develop probabilistic SR-q-DL to quantify the uncertainty of the network output. Wang et al. \cite{wang2024ultrafast} propose a novel fast diffusion tensor imaging method that requires only three diffusion directions per slice. High-quality DTI reconstruction can be achieved through the integration of multi-slice information and T1-weighted images. Although these methods effectively utilize information in the diffusion domain, they overlook spatial resolution, potentially resulting in limited spatial detail and structural information in the images, which may affect precise anatomical analysis\cite{yin2018joint}.

Although the previously mentioned methods improve dMRI data resolution, they are limited to addressing super-resolution in either the spatial or diffusion domain. Additionally, most methods use pixel-wise loss functions, which have limited capacity to recover image details. In this paper, we propose SARL-dMRI, a novel spatial-angular representation learning framework that integrates spatial and diffusion domain information for high-fidelity continuous super-resolution of diffusion MRI. Specifically, it leverages implicit neural representations\cite{chen2021learning} and spherical harmonics\cite{descoteaux2007regularized} to learn continuous representations in the spatial and diffusion domains, enhancing the spatial and angular resolution of dMRI data and improving the accuracy of downstream tasks. This results in more accurate microstructural parameter estimation, improved image quality, and enhanced anatomical accuracy. Our main contributions can be summarized as follows:

1) Novel Spatial-Angular Representation Learning Framework (SARL-dMRI): The development of the SARL-dMRI framework, which employs implicit neural representations and spherical harmonics, enables high-fidelity, continuous super-resolution in diffusion MRI, enhancing both spatial and angular resolution simultaneously.

2) Data-Fidelity Module and Wavelet-Based Frequency Loss Design: Introduction of a data-fidelity module and a wavelet-based frequency loss, which ensures the super-resolved dMRI images remain consistent with the original data and retain fine details, preserving image fidelity during resolution enhancement.

3) Superior Performance in Microstructural Estimation and Generalization: Extensive experiments demonstrate that SARL-dMRI significantly improves dMRI data resolution, enhances microstructural parameter estimation accuracy, and outperforms five state-of-the-art methods, offering better generalization and adaptability across varying downsampling rates.

The remainder of this paper is organized as follows. Section \ref{sec2} introduces the basic concepts of implicit neural representations and spherical harmonics. Section \ref{sec3} provides a detailed description of our method, SARL-dMRI. Section \ref{sec4} describes the experimental setup, including the dataset, comparison methods, and implementation details. Section \ref{sec5} presents and analyzes the experimental results. Section \ref{sec6} concludes the paper.

% Section 2 Preliminaries
\section{Preliminaries}\label{sec2}

% Section 2.1 Implicit Neural Representation Learning
\subsection{Implicit Neural Representations}\label{sec2.1}
Implicit neural representations (INRs) aim to use multilayer perceptrons (MLPs) to map coordinates to signal sources. This approach is widely used in medical imaging tasks \cite{molaei2023implicit}, such as image reconstruction\cite{shen2022nerp}, super-resolution \cite{wu2022arbitrary}, segmentation\cite{zhang2021nerd}, and others. Specifically, a neural network can represent an image as a continuous function. The neural network $\mathcal{F}_{\theta } $, parameterized by $\theta$, can be defined as:

% eq1
\begin{equation}
\label{eq1}
    I = \mathcal{F}_{\theta }(c), c\in [-1, 1]^{n}, I\in \mathbb{R} , 
\end{equation}
where the input $c$ is the normalized coordinate index in the image spatial field, and the output $I$ denotes the intensity value corresponding to the coordinate $c$. The neural network $\mathcal{F}_{\theta}$ learns the nonlinear relationship from coordinates to the intensity values in the image, effectively encoding the internal information of the entire image into the network parameters. Thus, the neural network $\mathcal{F}_{\theta}$ is also regarded as a neural representation of the image. In dMRI data, our coordinate indices are typically constructed based on the size of the dMRI data and normalized to a range of [-1, 1] along each dimension. For \eqref{eq1}, the neural network $\mathcal{F}_{\theta}$ performs super-resolution only on a single patient’s dMRI data and cannot generalize to data from multiple patients. In \ref{sec3.3.2}, we will describe a method to achieve super-resolution for dMRI data across various patients by integrating additional auxiliary information.

% Section 2.2 Spherical Harmonics Representation
\subsection{Spherical Harmonics Representation}\label{sec2.2}
In dMRI, the angular characteristics of the signal can be represented using spherical harmonics (SH)\cite{descoteaux2007regularized,anderson2005measurement,ha2022spharm}. Specifically, the dMRI signal can be approximately represented as a linear combination of spherical harmonics basis\cite{anderson2005measurement}, as follows:

% eq2
\begin{equation}
    \label{eq2}
    I(\theta ,\varphi )=\sum_{l=0}^{L} \sum_{m=-l}^{l}C_{l}^{m} Y_{l}^{m}(\theta ,\varphi) ,
\end{equation}
where $I(\theta, \varphi )$ denotes the dMRI signal, $\theta$ and $\varphi$ are the azimuthal and polar angles in the spherical coordinate system, used to describe the diffusion direction $(\theta \in \left [ 0, \pi \right ], \varphi \in \left [ 0, 2\pi \right ] )$, $l$ denotes the order and $m$ denotes the phase factor, $C_{l}^{m}$ denotes the SH coefficients. $Y_{l}^{m}$ denotes the SH and is a basis for complex functions on the unit sphere, which can be defined in a complex domain:

%eq3
\begin{equation}
    \label{eq3}
    Y_{l}^{m}(\theta ,\varphi)=\sqrt{\frac{2l+1}{4\pi}\frac{(l-m)!}{(l+m)!}  }P_{l}^{m}(\cos \theta )e^{im\varphi} ,
\end{equation}
where $P_{l}^{m}$ is an associated Legendre polynomial. Accordingly, the spherical harmonic coefficients can be calculated as:

%eq4
\begin{equation}
    \label{eq4}
    C_{l}^{m}=\int _{\Omega }I(\theta ,\varphi )Y_{l}^{m}(\theta ,\varphi )d\Omega ,
\end{equation}
where $\Omega$ denotes integration over the unit sphere. Once the spherical harmonic coefficients $C$ are obtained, the corresponding diffusion data can be computed for any given diffusion direction represented by $\theta$ and $\varphi$.

% Section 3 Methods
\section{Methods}\label{sec3}

% Section 3.1 Problem Definition
\subsection{Problem Definition}\label{sec3.1}
To improve the accuracy of microstructural parameter estimation, dMRI data with high spatial and angular resolution is typically required. This leads to longer scan times, increasing the demand for techniques that enhance both the spatial and angular resolution of dMRI data. For the problem of spatial-angular super-resolution (SASR) in dMRI, the process of generating a low-resolution (LR) diffusion image $I_{lr}$ ($H_{1} \times W_{1} \times Z_{1} \times N_{1}$) can be defined as:

%eq5
\begin{equation}
\label{eq5}
    I_{lr}=D\mathcal{Q}I_{hr}+\eta ,   
\end{equation}
where $\mathcal{Q}$ denotes the sampling operator in the diffusion wavevector domain (q-space), $D$ denotes the downsampling operator in the spatial domain (x-space), $\eta  $ is the corresponding acquisition noise. $I_{hr}$ ($H_{2} \times W_{2} \times Z_{2} \times N_{2}$) represents the high-resolution (HR) diffusion image. $H$, $W$, and $Z$ represent the spatial dimensions, and $N$ represents the number of diffusion directions. SASR aims to recover the HR image $I_{hr}$ from the LR image $I_{lr}$. Due to factors such as down-sampling operations and system noise, this recovery process is considered an ill-posed inverse problem. Prior knowledge is typically introduced to better alleviate this issue. Consequently, the solving process can be transformed into the following optimization problem:

%eq6
\begin{equation}
\label{eq6}
   \hat{I}_{hr}= arg \min_{I_{hr}}\{ \frac{1}{2} \left \| D\mathcal{Q}I_{hr} -I_{lr} \right \| ^{2}_{2} +\lambda \mathfrak{R}(I_{hr})  \} ,
\end{equation}
where the first term is the data fidelity term, which constrains the difference between the degraded HR image $I_{hr}$ and the observed LR image $I_{lr}$. The second term is the regularization term, which alleviates the ill-posed problem by introducing prior information. The weight $\lambda$ is used to balance the contributions of the two terms. In this paper, the regularization term is implemented using a neural network.

% fig1
\begin{figure*}[htbp]
    \centering
    \includegraphics[width=\linewidth]{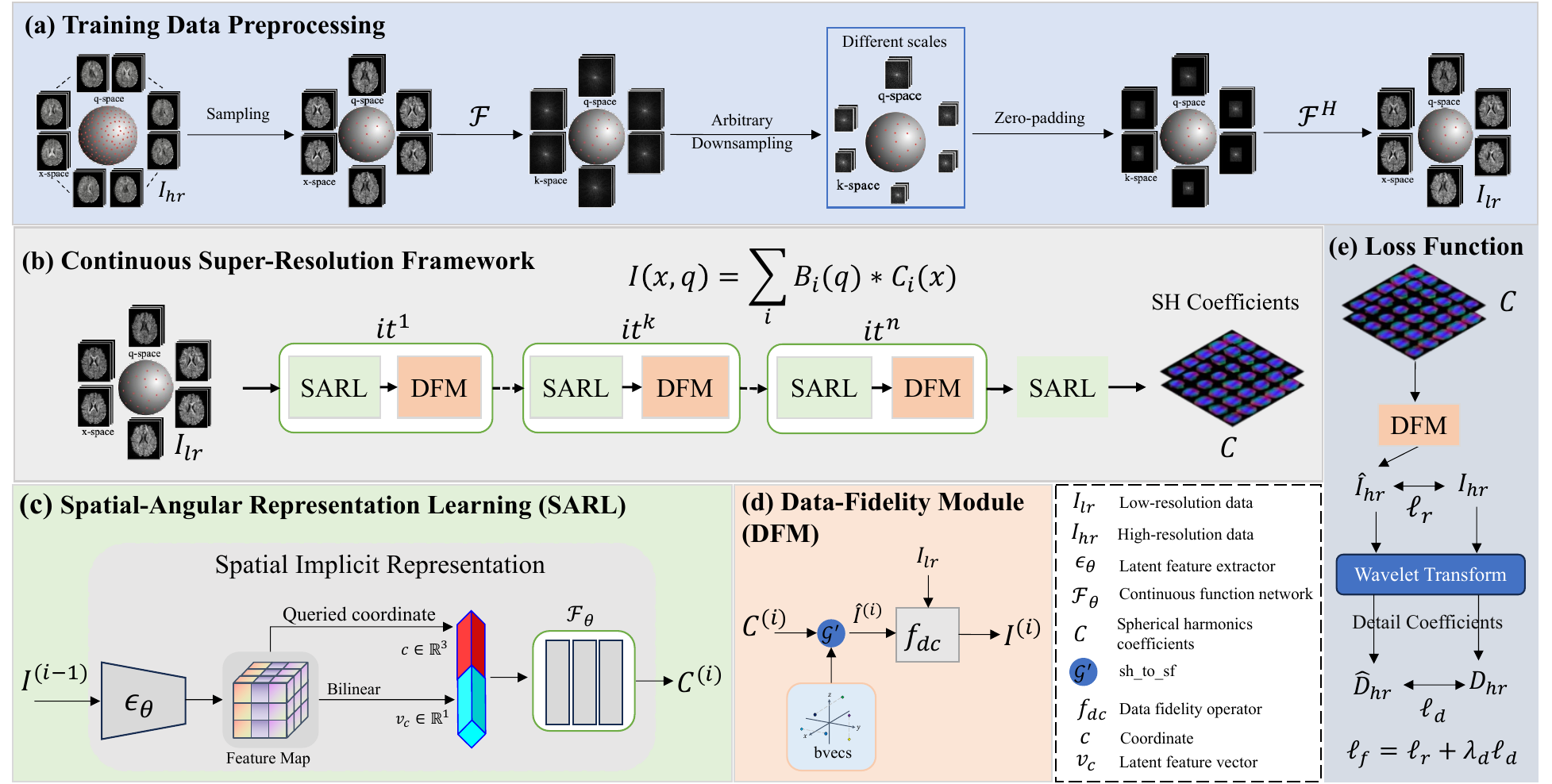}
    \caption{The overview of SARL-dMRI method. (a) Training data preprocessing. The low-resolution image $I_{lr}$ is obtained by simultaneously downsampling in both q-space and x-space. (b) Continuous super-resolution framework. It is an end-to-end framework with alternating iterations of the SARL and DFM modules. (c) Spatial-angular representation learning (SARL). The dMRI signal is modeled continuously using implicit neural representations and spherical harmonics. (d) Data-fidelity module (DFM). The data fidelity operator ensures that the super-resolved image does not deviate from the original input image. (e) Loss function. The anatomical structures of the image are preserved by incorporating wavelet-based frequency loss.}
    \label{fig1}
\end{figure*}

% Section 3.2 Continuous Super-Resolution Framework
\subsection{Continuous Super-Resolution Framework}\label{sec3.2}

We design a continuous super-resolution framework to achieve continuous super-resolution in diffusion MRI, as shown in Fig. \ref{fig1}(b). By decomposing \eqref{eq6}, we construct an end-to-end super-resolution framework that operates through alternating iterations. It consists of two parts: a spatial-angular representation learning (SARL) module and a data fidelity module (DFM). SARL aims to generate HR spherical harmonic coefficients from low spatial-angular resolution dMRI data, with details provided in \ref{sec3.3}. The DFM ensures that super-resolved images do not deviate from the original input data, as described in \ref{sec3.4}. The input to the entire framework is the low spatial-angular resolution image $I_{lr}\in \mathbb{R}^{H_{1}\times W_{1} \times 3 \times N_{1}}$, and its generation process is illustrated in Fig. \ref{fig1}(a). To balance computational efficiency and performance, we use three neighboring slices to predict the middle slice in this paper, leveraging complementary information from adjacent slices while reducing the demand for computational resources. First, the high-resolution image $I_{hr}\in \mathbb{R}^{H_{2}\times W_{2} \times 3 \times N_{2}}$ is sampled in q-space, then transformed into the Fourier domain for arbitrary scale k-space downsampling, followed by zero-filling. Finally, the zero-padded k-space data is transformed into the image domain to obtain the low-resolution input data $I_{lr}\in \mathbb{R}^{H_{1}\times W_{1} \times 3 \times N_{1}}$. For the continuous super-resolution framework, $I_{lr}$ passes through $n$ alternating iterations of the SARL and DFM modules, followed by an additional SARL module to produce the model's output $C\in \mathbb{R}^{H_{2}\times W_{2} \times 1 \times 28}$, where $C$ denotes the spherical harmonic coefficients. As recommended in \cite{descoteaux2007regularized}, we adopt spherical harmonic coefficients of order 6, resulting in 28 coefficients. This ensures that the representation of the dMRI signal remains accurate and stable, avoiding the issues caused by choosing an excessively high or low order. Finally, the corresponding diffusion image can be obtained based on \eqref{eq2} for any given diffusion direction. These reconstructed high-resolution diffusion images enable the estimation of microstructural parameter maps for downstream tasks, such as FA, MD, diffusion tensors, and fibre orientation distribution (FOD) \cite{tournier2007robust}.

%3.3
\subsection{Spatial-Angular Representation Learning}\label{sec3.3}
The primary goal of the SARL module is to generate HR spherical harmonic coefficients from LR diffusion images, as illustrated in Fig. \ref{fig1}(c). Specifically, it consists of two components: a latent feature extractor $\epsilon_{\theta}$ and a continuous function network $\mathcal{F}_{\theta}$. The latent feature extractor extracts features from low-resolution dMRI data, while the continuous function network predicts SH coefficients at any given coordinate by combining the coordinate with the corresponding latent feature vectors as input. The details of the two components will be described in the following sections.

%3.3.1
\subsubsection{Latent Feature Extractor}\label{sec3.3.1}
Inspired by \cite{wu2022arbitrary,chen2021learning}, we use a Residual Dense Network (RDN) \cite{zhang2018residual} as our feature extractor to derive latent feature vectors from low-resolution dMRI data. To ensure the extracted feature maps maintain their original dimensions, we removed the final upsampling operation from the original RDN network and substituted all 2D convolutional layers with 3D convolutional layers. The number of output channels of the network is consistent with the number of input diffusion directions. The latent feature extractor $\epsilon_{\theta}$ takes $I_{lr}$ as input and outputs the feature maps
$V_{lr}$. Bilinear interpolation is then employed to obtain the latent feature vector $v_{c}$, corresponding to the HR image coordinates $c$. The coordinate $c$ and latent feature vector $v_{c}$ are then input into the continuous function network $\mathcal{F}_{\theta}$ to obtain the spherical harmonic coefficients at the corresponding coordinate $c$. 

%3.3.2
\subsubsection{Continuous Function Network}\label{sec3.3.2}
The continuous function network $\mathcal{F}_{\theta}$ consists of eight standard fully connected (FC) layers, with a ReLU activation function following each FC layer except for the last one. In addition, to reduce information loss, we add a residual connection between the network’s input and the output of the fourth ReLU activation function. The goal of $\mathcal{F}_{\theta}$ is to predict the spherical harmonic coefficients at any spatial coordinate $c$. The specific process is as follows:

%eq7
\begin{equation}
    \label{eq7}
    C =\mathcal{F}_{\theta}(c, v_{c}) ,
\end{equation}
where $C$ denotes the spherical harmonic coefficients corresponding to coordinate $c$, and $v_{c}$ denotes the latent feature vector at coordinate $c$. Instead of using spatial coordinates alone as input to the continuous function network, we combine the spatial coordinates with the corresponding latent feature vector. This approach effectively integrates local semantic information from the low-resolution data, enhancing the ability of $\mathcal{F}_{\theta}$ to recover image details while enabling the super-resolution of dMRI data from multiple patients.

%3.4
\subsection{Data-Fidelity Module}\label{sec3.4}
\begin{figure}[htbp]
    \centering
    \includegraphics[width=\linewidth]{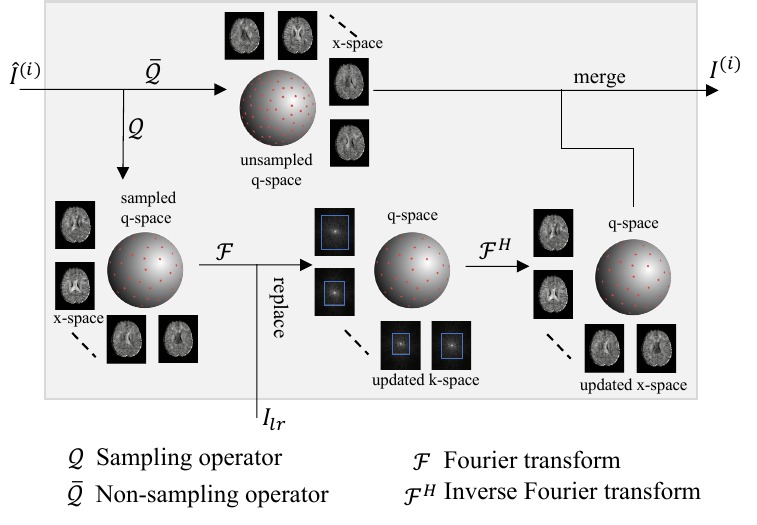}
    \caption{The data fidelity operator. $\mathcal{Q}$ denotes the sampling operator, while $\bar{\mathcal{Q}}$ represents the non-sampling operator in q-space. $\mathcal{F}$ denotes the Fourier transform, and $\mathcal{F}^{H}$ denotes the inverse Fourier transform.}
    \label{fig2}
\end{figure}

To ensure that the overall structure and contrast of the super-resolved dMRI data do not deviate, we design a data fidelity module (DFM), as illustrated in Fig. \ref{fig1}(d). The specific process is as follows: First, the SH coefficients $C^{(i)}$ obtained from the SARL module are used to generate the diffusion-weighted (DW) images $\hat{I}^{(i)}$ for given diffusion direction. Then, the super-resolved DW images $\hat{I}^{(i)}$ are sampled using $\mathcal{Q}$, and both it and the original low-resolution data $I_{lr}$ are input into the 
 data fidelity operator $f_{df}$ (Fig. \ref{fig2}) to obtain the updated high-resolution DW images $I^{(i)}$. By introducing a data fidelity operator, the reliability of the super-resolved image can be enhanced.

%fig2

%3.5
\subsection{Wavelet-based Frequency Loss}\label{sec3.5}
The pixel-wise loss function has been widely used in image super-resolution tasks\cite{zhao2016loss}. While pixel-wise loss effectively minimizes overall pixel differences, it inherently limits the ability to preserve fine image details, such as edges and textures. The wavelet transform can capture high-frequency information across different scales of an image, thus facilitating the learning of its multi-scale structure\cite{guo2017deep}. To this end, we introduce a wavelet-based frequency loss, as shown in Fig. \ref{fig1}(e). The specific calculation process is as follows:

%eq8
\begin{equation}
\label{eq8}
    \ell _{d}=\sum_{l=1}^{L}\left \| \hat{D}_{hr}^{l}-D_{hr}^{l}  \right \| _{1} ,
\end{equation}
where $\hat{D}_{hr}^{l}$ and $D_{hr}^{l}$ represent the detail coefficients obtained from the super-resolved image and the reference image via wavelet transformation, respectively, where $l$ denotes the number of wavelet decomposition levels. This paper uses Daubechies (Db) 4 wavelet decomposition with 3 levels. Multi-level wavelet decomposition can more effectively capture the high-frequency information of an image, thereby improving the detail recovery of dMRI data. The total loss function is:

%eq9
\begin{equation}
    \label{eq9}
    \ell _f = \ell_r+\lambda_d\ell_d  ,
\end{equation}
where $\ell_{r}$ denotes the $l$1 loss between $\hat{I}_{hr}$ and $I_{hr}$, $\lambda_{d}$ is used to adjust the weight of $\ell_{d}$.

% Section 4 Experiments
\section{Experiments}\label{sec4}
In this section, we introduce the dataset used in the experiments, followed by the comparison methods, the experimental setup, and the evaluation metrics.

% Section 4.1 Dataset
\subsection{Datasets}\label{sec4.1}
The Human Connectome Project (HCP) Young Adult 3T dataset was used to validate the effectiveness of our method, and it was obtained using a standard 32-channel Siemens receive head coil \cite{glasser2013minimal}. Each gradient table includes approximately 270 diffusion directions and 18 b=0 acquisitions. The diffusion image consists of three shells with b-values of 1000, 2000, and 3000 $s/mm^{2}$, each shell typically contains around 90 diffusion directions. The data resolution is isotropic at 1.25 mm. Data from 50 subjects with b=1000 $s/mm^{2}$ were used for this study. All data underwent the HCP preprocessing pipeline, with 35 subjects used for training, 5 for validation, and 10 for testing.

To obtain the training data, we first normalized all data by dividing it by the mean of the b=0 DW images. Then, the subject data is divided into multiple samples, each consisting of three consecutive slices along the slice direction, with 75 samples extracted from each subject. To generate the reference DTI maps, diffusion tensor fitting was performed on the b=1000 $s/mm^{2}$ and b=0 data using the DIPY \footnote{https://dipy.org/} toolkit, yielding the ground truth FA and MD parameter maps and diffusion tensors for testing. In addition, we compute the FOD using the method proposed in \cite{tournier2007robust} and visualize it using MRtrix3 \footnote{https://www.mrtrix.org/}.

% table 1
\begin{table*}[htbp]
\centering
\caption{Quantitative comparisons of high-resolution diffusion images from different methods are shown across four in-distribution scenarios. Bold numbers indicate the best results. }
\label{tab1}
\resizebox{0.95\textwidth}{!}{
\tiny % 控制字体大小
\begin{tabular}{ccccccc}
\toprule[1pt]
\multirow{2}{*}{\begin{tabular}[c]{@{}c@{}}Method\\ (Spatial+Angular)\end{tabular}} & \multicolumn{3}{c}{S$\times$2, Q$\times$15}             & \multicolumn{3}{c}{S$\times$2, Q$\times$3}                \\ \cline{2-7} 
& PSNR/dB        & SSIM          & NRMSE         & PSNR/dB        & SSIM          & NRMSE         \\ \midrule[1pt]
ShuffleUNet+SH                                                                      & 28.5559±0.6169 & 0.9474±0.0070 & 0.1774±0.0099 & 29.4851±0.6148 & 0.9599±0.0058 & 0.1586±0.0091 \\ %\hline
ResCNN+SH                                                                           & 28.4279±0.6103 & 0.9473±0.0070 & 0.1805±0.0101 & 29.9364±0.5838 & 0.9643±0.0052 & 0.1523±0.0087 \\ %\hline
ShuffleUNet+RCNN                                                                    & 29.0731±0.6365 & 0.9550±0.0066 & 0.1677±0.0101 & 29.6149±0.5443 & 0.9614±0.0053 & 0.1560±0.0070 \\ %\hline
ResCNN+RCNN                                                                         & 29.0009±0.6232 & 0.9552±0.0066 & 0.1696±0.0010 & 30.1720±0.5023 & 0.9665±0.0044 & 0.1475±0.0062 \\ %\hline
% ShuffleUNet+3D U-net

% &
% 28.7858±0.6092&
% 0.9492±0.0064&
% 0.1733±0.0096&
% 29.5185±0.5011&
% 0.9587±0.0051&
% 0.1581±0.0052 \\
ArSSR+SH                                                                            & 28.3245±0.6066 & 0.9468±0.0071 & 0.1829±0.0101 & 29.6652±0.6164 & 0.9632±0.0055 & 0.1569±0.0097 \\ %\hline
SARL-dMRI                                                                           & \textbf{29.7756±0.6584} & \textbf{0.9603±0.0061} & \textbf{0.1542±0.0097} & \textbf{30.9555±0.5508} & \textbf{0.9694±0.0043} & \textbf{0.1358±0.0066} \\ \midrule[1pt]
\multirow{2}{*}{\begin{tabular}[c]{@{}c@{}}Method\\ (Spatial+Angular)\end{tabular}} & \multicolumn{3}{c}{S$\times$3, Q$\times$15}             & \multicolumn{3}{c}{S$\times$3, Q$\times$3}                \\ \cline{2-7} 
& PSNR/dB        & SSIM          & NRMSE         & PSNR/dB        & SSIM          & NRMSE         \\ \midrule[1pt]
ShuffleUNet+SH                                                                      & 27.3543±0.5253 & 0.9318±0.0076 & 0.2038±0.0085 & 27.9052±0.5228 & 0.9434±0.0067 & 0.1923±0.0082 \\ %\hline
ResCNN+SH                                                                           & 27.1195±0.5200 & 0.9313±0.0076 & 0.2108±0.0085 & 27.8069±0.4851 & 0.9457±0.0062 & 0.1939±0.0074 \\ %\hline
ShuffleUNet+RCNN                                                                    & 27.5448±0.5212 & 0.9382±0.0073 & 0.1995±0.0089 & 27.9656±0.4841 & 0.9447±0.0064 & 0.1908±0.0071 \\ %\hline
ResCNN+RCNN                                                                         & 27.3633±0.5122 & 0.9382±0.0072 & 0.2053±0.0086 & 27.9255±0.4340 & 0.9477±0.0056 & 0.1907±0.0058 \\ %\hline
% ShuffleUNet+3D U-net

% &
% 27.4360±0.5156&
% 0.9331±0.0073&
% 0.2023±0.0083&
% 27.8965±0.4573&
% 0.9419±0.0063&
% 0.1929±0.0056 \\
ArSSR+SH                                                                            & 27.1743±0.5111 & 0.9324±0.0076 & 0.2097±0.0083 & 27.8732±0.5076 & 0.9470±0.0062 & 0.1928±0.0082 \\ %\hline
SARL-dMRI                                                                           & \textbf{28.9717±0.6192} & \textbf{0.9535±0.0066} & \textbf{0.1690±0.0095} & \textbf{29.6860±0.5410} & \textbf{0.9610±0.0051} & \textbf{0.1557±0.0071} \\ \bottomrule[1pt]
\multicolumn{7}{p{251pt}}{S: Spatial downsampling factor, S$\times$ 2: 2.5mm$\to$1.25mm, S$\times$ 3: 3.75mm$\to$1.25mm.} \\ 
\multicolumn{7}{p{251pt}}{Q: Angular sampling factor, Q$\times$15: 6 directions$\to$90 directions, Q$\times$3: 30 directions$\to$90 directions.}
\end{tabular}
}
\end{table*}

% Section 4.2 Baseline methods
\subsection{Comparison methods}\label{sec4.2}
Existing deep learning super-resolution methods for diffusion data predominantly focus on a single aspect, enhancing spatial or angular resolution. Therefore, we adopt five hybrid super-resolution methods as our comparison method. In the spatial domain, super-resolution includes two methods that learn fixed scaling factors: ShuffleUNet \cite{chatterjee2021shuffleunet} and ResCNN \cite{du2020super}, as well as one method that learns continuous scaling factors: ArSSR \cite{wu2022arbitrary}. Super-resolution methods in the diffusion domain include SH \cite{descoteaux2007regularized} and deep learning-based angular super-resolution method: RCNN \cite{lyon2022angular}. The five hybrid comparison methods are denoted as ShuffleUNet+SH, ResCNN+SH, ShuffleUNet+RCNN, ResCNN+RCNN, and ArSSR+SH. The order of the SH is set to 6, and ArSSR learns the mapping relationship between 2 and 3.

% fig 3
\begin{figure}[htbp]
    \centering
\includegraphics[width=\linewidth]{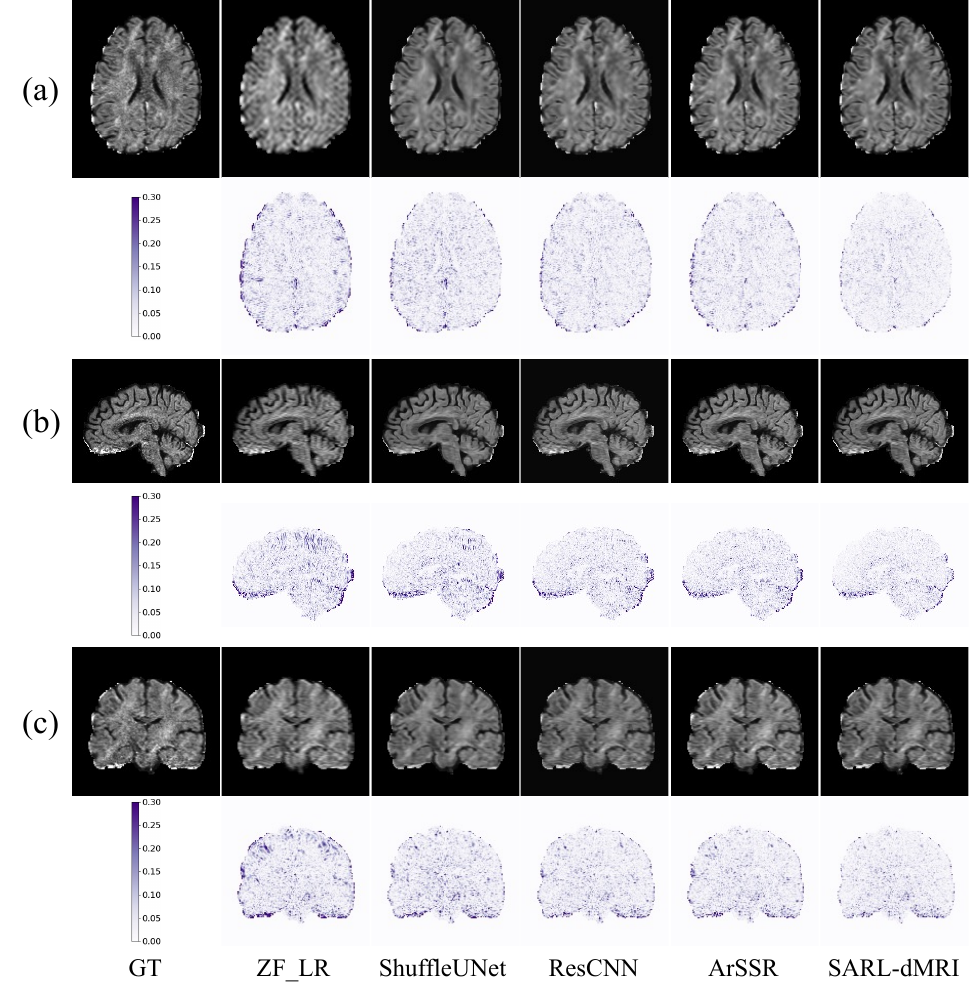}
    \caption{The visualization results of high-resolution diffusion images obtained using different spatial-domain super-resolution methods are presented under in-distribution scenario (S$\times$3: 3.75mm$\to$1.25mm, Q$\times$3: 30 directions$\to$90 directions), including three views: (a) axial, (b) sagittal, and (c) coronal. Rows 2, 4, and 6 correspond to the respective error maps. ZF\_LR denotes a zero-filled low-resolution image.}
    \label{fig3}
\end{figure}

% Section 4.3 Implementation Details
\subsection{Implementation Details}\label{sec4.3}
All the experiments were conducted using the PyTorch framework with two NVIDIA RTX A6000 GPUs (each with 48GB memory). The model is trained with the AdamW optimizer for 200 epochs, starting with an initial learning rate of $1e^{-4}$, which is reduced to $1e^{-5}$ after 100 epochs. During training, the model learns continuous scaling factors within the range [2, 3]. Model performance is evaluated using common quantitative metrics, including peak signal-to-noise ratio (PSNR), structural similarity index (SSIM), and normalized root mean square error (NRMSE). The code will be available at https://github.com/ternencewu123/SARL-dMRI.

% Section 5 Results
\section{Results}\label{sec5}
To fully validate the performance of our method, we conducted experiments in two main areas. This first evaluated dMRI reconstruction performance, including both in-distribution and out-of-distribution scenarios. The second assessed its effectiveness in improving downstream dMRI tasks, including FA, MD, tensor, and FOD \cite{tournier2007robust}.

%table 2
\begin{table}[ht]
\centering
\caption{The quantitative results of different methods under out-of-distribution spatial downsampling factors. Bold numbers indicate the best results. }
\label{tab2}
\renewcommand{\arraystretch}{1.5} % 调整行距
\resizebox{0.5\textwidth}{!}{
% \tiny % 控制字体大小
\begin{tabular}{ccccc}
\toprule[1pt]
\multirow{2}{*}{\begin{tabular}[c]{@{}c@{}}Method\\ (Spatial+Angular)\end{tabular}} & \multicolumn{2}{c}{S $\times$ 3.2, Q $\times$ 3} & \multicolumn{2}{c}{S $\times$ 3.6, Q $\times$ 3} \\ \cline{2-5} 
 & PSNR/dB        & SSIM          & PSNR/dB        & SSIM          \\ \midrule[1pt]
ShuffleUNet+SH                                                                      & 27.5833±0.4971 & 0.9385±0.0067 & 26.6000±0.4398 & 0.9248±0.0069 \\ 
ResCNN+SH                                                                           & 27.4054±0.4618 & 0.9399±0.0063 & 26.4136±0.4285 & 0.9240±0.0068 \\ 
ShuffleUNet+RCNN                                                                    & 18.4680±0.4211 & 0.7864±0.0200 & 18.1200±0.4048 & 0.7810±0.0202 \\ 
ResCNN+RCNN                                                                         & 18.4950±0.4156 & 0.7875±0.0199 & 18.0911±0.4126 & 0.7806±0.0204 \\ 
ArSSR+SH                                                                            & 27.3718±0.4876 & 0.9404±0.0065 & 26.3920±0.4478 & 0.9246±0.0070 \\ 
SARL-dMRI                                                                           & \textbf{29.2456±0.5279} & \textbf{0.9576±0.0054} & \textbf{28.1444±0.4442} & \textbf{0.9483±0.0057} 
\\ \bottomrule[1pt]
\multicolumn{5}{p{251pt}}{S: Spatial downsampling factor, S$\times$3.2: 4mm$\to$1.25mm, S$\times$3.6: 4.5mm$\to$1.25mm.}\\
\multicolumn{5}{p{251pt}}{Q: Angular sampling factor, Q$\times$3: 30 directions$\to$90 directions.}
\end{tabular}
}
\end{table}

%table3
\begin{table*}[ht]
\centering
\caption{Quantitative results of FA and MD parameter maps and diffusion tensors obtained by different methods. Bold numbers indicate the best results. }
\label{tab3}
\renewcommand{\arraystretch}{1.5} % 调整行距
\resizebox{\textwidth}{!}{
\tiny % 控制字体大小
\begin{tabular}{cccccccccc}
% \hline
\toprule[1pt]
\multirow{3}{*}{\begin{tabular}[c]{@{}c@{}}Method\\ (Spatial+Angular)\end{tabular}} & \multicolumn{9}{c}{S $\times$ 3, Q $\times$ 3}                                                                                                                   \\ \cline{2-10} 
& \multicolumn{3}{c}{FA}                         & \multicolumn{3}{c}{MD}                         & \multicolumn{3}{c}{Tensors}                   \\ \cline{2-10} 
& PSNR/dB        & SSIM          & NRMSE         & PSNR/dB        & SSIM          & NRMSE         & PSNR/dB        & SSIM          & NRMSE         \\ \midrule[1pt]
ShuffleUNet+SH                                                                      & 27.8169±0.4660 & 0.9328±0.0048 & 0.2844±0.0114 & 33.2160±4.3894 & 0.9473±0.0384 & 0.4109±0.1621 & 31.9301±4.2307 & 0.9776±0.0118 & 0.2188±0.1041 \\ %\hline
ResCNN+SH                                                                           & 28.1148±0.5094 & 0.9379±0.0049 & 0.2749±0.0127 & 33.7563±4.6487 & 0.9449±0.0393 & 0.3776±0.1207 & 33.5193±3.4384 & 0.9783±0.0112 & 0.1854±0.0995 \\ %\hline
ShuffleUNet+RCNN                                                                    & 27.6850±0.4874 & 0.9265±0.0059 & 0.2887±0.0108 & 32.0081±3.6464 & 0.9442±0.0326 & 0.4656±0.1685 & 33.1907±3.6910 & 0.9811±0.0088 & 0.1880±0.0943 \\ %\hline
ResCNN+RCNN                                                                         & 27.9229±0.5202 & 0.9303±0.0055 & 0.2810±0.0116 & 33.8246±4.6153 & 0.9471±0.0383 & 0.3839±0.1530 & 34.0630±6.0642 & 0.9791±0.0110 & 0.1940±0.1187 \\ %\hline
ArSSR+SH                                                                            & 28.3508±0.5088 & 0.9450±0.0046 & 0.2675±0.0125 & 32.0393±3.8590 & 0.9376±0.0368 & 0.4549±0.1206 & 33.1275±5.2331 & 0.9763±0.0109 & 0.2083±0.1198 \\ %\hline
SARL-dMRI                                                                           & \textbf{29.0966±0.5110} & \textbf{0.9474±0.0038} & \textbf{0.2455±0.0112} & \textbf{35.1876±4.7325} & \textbf{0.9657±0.0296} & \textbf{0.3481±0.1953} & \textbf{36.1412±5.2992} & \textbf{0.9859±0.0121} & \textbf{0.1415±0.0830} \\ \bottomrule[1pt]
\multicolumn{10}{p{251pt}}{S: Spatial downsampling factor, S$\times$3: 3.75mm$\to$1.25mm. Q: Angular sampling factor, Q$\times$3: 30 directions$\to$90 directions.}
\end{tabular}
}
\end{table*}

% figure 4
\begin{figure*}[htbp]
    \centering
    \includegraphics[width=0.95\linewidth]{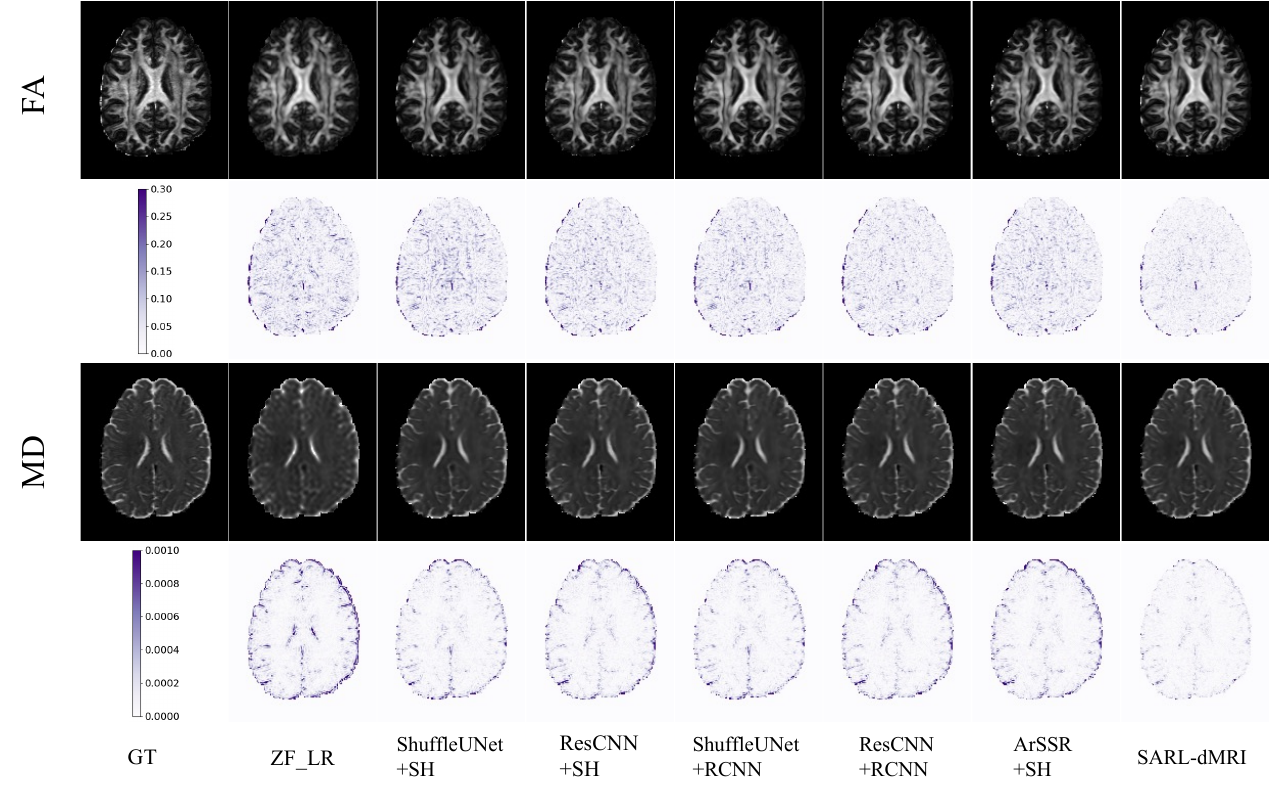}
    \caption{Qualitative results of FA and MD parameter maps obtained by different methods. Rows 1 and 3 present the visualizations of FA and MD, respectively, while rows 2 and 4 present the corresponding error maps. ZF\_LR denotes a zero-filled low-resolution image.}
    \label{fig4}
\end{figure*}

% Section 5.1 Comparison of HR Diffusion Image Reconstruction
\subsection{Comparison Results of HR Diffusion Image}\label{sec5.1}

%Section 5.1.1 In-distribution Scenarios
\subsubsection{In-distribution Scenarios}\label{sec5.1.1}
Table \ref{tab1} shows the quantitative results of super-resolution diffusion images obtained by different methods under four in-distribution scenarios. From a vertical perspective, our method achieves the best performance across all spatial downsampling rates when the angular sampling rate is fixed. Similarly, from a horizontal perspective, our method consistently achieves the best results across different angular sampling rates when the spatial downsampling rate is fixed. Notably, even at an overall downsampling factor of 45$\times$, our method achieves PSNR, SSIM, and NRMSE scores of 28.9717 $\pm$ 0.6192, 0.9535 $\pm$ 0.0066, and 0.1690 $\pm$ 0.0095, respectively, highlighting the effectiveness of our spatial-angular representation learning framework. Moreover, the experimental results indicate that single-stage super-resolution methods outperform two-stage methods by simultaneously learning representation from two domains. 

Fig. \ref{fig3} shows the visualization results of high-resolution diffusion images obtained using different spatial domain super-resolution methods under an in-distribution scenario (S$\times$3: 3.75mm$\to$1.25mm, Q$\times$3: 30 directions$\to$90 directions), including three views: axial, sagittal, and coronal. The visualization results and error maps demonstrate that our method outperforms others and restores more image details. Overall, the proposed single-stage super-resolution method outperforms the two-stage super-resolution method. Even under higher downsampling factors, our method obtains superior image quality, further validating its effectiveness.

%fig 4
% \begin{figure}[htbp]
%     \centering
%     \includegraphics[width=\linewidth]{LaTeX/figures/radar1.pdf}
%     \caption{The quantitative results of different methods under out-of-distribution spatial downsampling factors (S$\times$3.2: 4mm$\to$1.25mm, S$\times$3.4: 4.25mm$\to$1.25mm, S$\times$3.6: 4.5mm$\to$1.25mm, S$\times$3.8: 4.75mm$\to$1.25mm, S$\times$4: 5mm$\to$1.25mm; Q$\times$3: 30$\to$90). The different axes of the radar chart represent different spatial downsampling factors.}
%     \label{fig4}
% \end{figure}

% Section 5.1.2 Out-of-distribution Scenarios
\subsubsection{Out-of-distribution Scenarios}\label{sec5.1.2}
To further validate the performance of our method in improving the quality of diffusion images, we conduct experiments under two out-of-distribution spatial downsampling factor scenarios. Among the methods, ShuffleUNet+SH, ResCNN+SH, ShuffleUNet+RCNN, and ResCNN+RCNN were evaluated using models trained in the in-distribution setting (S$\times$3, Q$\times$3), whereas ArSSR+SH and our method utilize models trained under the configuration (S: [2, 3], Q$\times$3). Table \ref{tab2} illustrates the quantitative results of different methods under two out-of-distribution scenarios (S$\times$3.2, Q$\times$3; S$\times$3.6, Q$\times$3). The experimental results indicate that our method outperforms the comparison methods under two scenarios, achieving superior performance. As shown in Table \ref{tab2}, the traditional angular super-resolution method SH shows lower sensitivity to spatial domain variations.
In contrast, deep learning-based angular super-resolution methods are more sensitive to such variations, which can negatively impact their performance. For example, a clear performance difference exists between traditional angular super-resolution methods (such as SH) and deep learning-based methods (such as RCNN) under out-of-distribution scenarios. Specifically, SH’s super-resolution performance is less affected by spatial domain variations caused by different downsampling factors, while RCNN shows a significant performance decline in these scenarios. Traditional methods have their advantages, particularly in terms of model stability. By employing implicit neural representations and spherical harmonics to learn continuous representations across both domains, we can more effectively adapt to out-of-distribution scenarios, thereby broadening the applicability of our method.

% figure 5
\begin{figure}[htbp]
    \centering
    \includegraphics[width=0.95\linewidth]{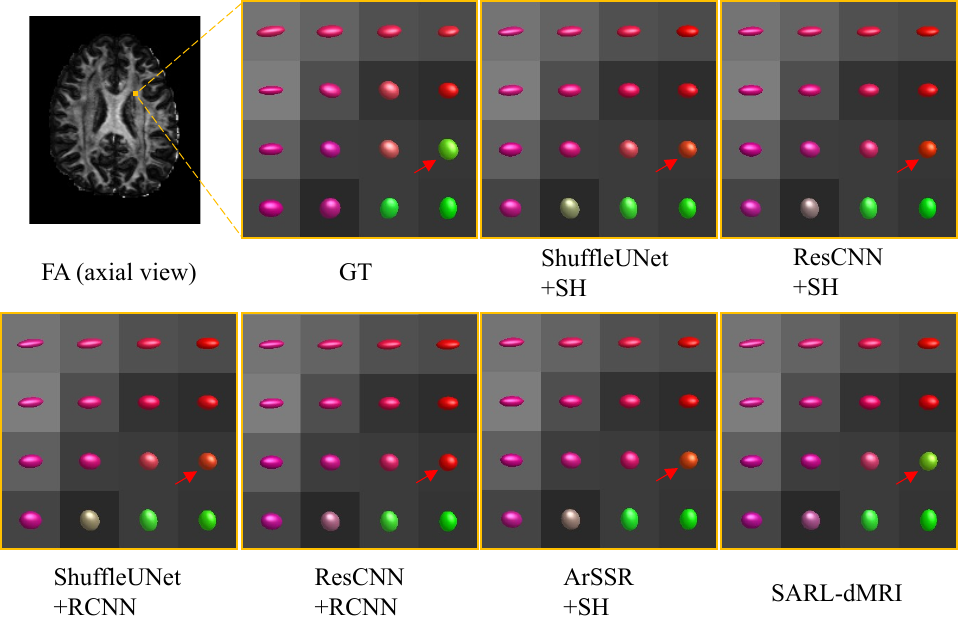}
    \caption{The visualized results of diffusion tensors derived from various methods. FA measure is used as the background for diffusion tensors visualization. The yellow rectangular box represents the locally magnified region.}
    \label{fig5}
\end{figure}

%table 4
\begin{table*}[!ht]
\centering
\caption{The alation studies of the data fidelity module and wavelet-based frequency loss $\ell_{d}$ were conducted under four different in-distribution scenarios. Bold numbers indicate the best results. }
\label{tab4}
\resizebox{0.95\textwidth}{!}{
\tiny % 控制字体大小
\begin{tabular}{ccccccc}
\toprule[1pt]
\multirow{2}{*}{Method} & \multicolumn{3}{c}{S $\times$ 2, Q $\times$ 15}                & \multicolumn{3}{c}{S $\times$ 2, Q $\times$ 3}                  \\ \cline{2-7} 
& PSNR/dB        & SSIM          & NRMSE         & PSNR/dB        & SSIM          & NRMSE         \\ \midrule[1pt]
Baseline                & 29.3385±0.6643 & 0.9573±0.0067 & 0.1621±0.0107 & 29.6705±0.6574 & 0.9616±0.0060 & 0.1556±0.0107 \\ %\hline
w/ DFM                     & 29.6931±0.6572 & 0.9599±0.0062 & 0.1557±0.0097 & 30.8920±0.5519 & 0.9693±0.0044 & 0.1363±0.0067 \\ %\hline
w/ $\ell_{d}$                      & 29.4316±0.6867 & 0.9578±0.0067 & 0.1605±0.0116 & 29.8505±0.6887 & 0.9623±0.0060 & 0.1529±0.0118 \\ %\hline
SARL-dMRI               & \textbf{29.7756±0.6584} & \textbf{0.9603±0.0061} & \textbf{0.1542±0.0097} & \textbf{30.9555±0.5508} & \textbf{0.9694±0.0043} & \textbf{0.1358±0.0066} \\ \midrule[1pt]
\multirow{2}{*}{Method} & \multicolumn{3}{c}{S $\times$ 3, Q $\times$ 15}                 & \multicolumn{3}{c}{S $\times$ 3, Q $\times$ 3}                  \\ \cline{2-7} 
& PSNR/dB        & SSIM          & NRMSE         & PSNR/dB        & SSIM          & NRMSE         \\ \midrule[1pt]
Baseline                & 28.4203±0.5977 & 0.9486±0.0071 & 0.1809±0.0102 & 28.7008±0.5865 & 0.9529±0.0065 & 0.1746±0.0102 \\ %\hline
w/ DFM                     & 28.8903±0.6137 & 0.9531±0.0066 & 0.1706±0.0095 & 29.6069±0.5444 & 0.9608±0.0052 & 0.1567±0.0072 \\ %\hline
w/ $\ell_{d}$                      & 28.5146±0.6086 & 0.9493±0.0072 & 0.1793±0.0110 & 28.8690±0.6222 & 0.9537±0.0065 & 0.1722±0.0119 \\ %\hline
SARL-dMRI               & \textbf{28.9717±0.6192} & \textbf{0.9535±0.0066} & \textbf{0.1690±0.0095} & \textbf{29.6860±0.5410} & \textbf{0.9610±0.0051} & \textbf{0.1557±0.0071} \\ \bottomrule[1pt]
\multicolumn{7}{p{251pt}}{S: Spatial downsampling factor, S$\times$2: 2.5mm$\to$1.25mm, S$\times$3: 3.75mm$\to$1.25mm.} \\ 
\multicolumn{7}{p{251pt}}{Q: Angular sampling factor, Q$\times$15: 6 directions$\to$90 directions, Q$\times$3: 30 directions$\to$90 directions.}
\end{tabular}
}
\end{table*}

% Section 5.2 Comparison results of downstream tasks
\subsection{Comparison results of downstream tasks}\label{sec5.2}
To further demonstrate the effectiveness of our method, we compare the performance of different methods on downstream tasks, including the generation of DTI parameter maps and the visualization of FOD.

% Section 5.2.1 DTI
\subsubsection{Comparison of DTI parameter maps}\label{sec5.2.1}

 Table \ref{tab3} presents the quantitative results of FA and MD parameter maps and diffusion tensors derived from different methods. The experimental results show that our proposed method, SARL-dMRI, outperforms other methods across all three metrics, with a notable advantage in the MD parameter map. Fig. \ref{fig4} visualizes the FA and MD parameter maps generated by different methods. The error maps demonstrate that the parameter maps generated by our method exhibit superior quality compared to those obtained by other methods. Moreover, the parameter maps generated by our method show better performance in detail preservation and noise suppression, indicating that the single-stage super-resolution method offers an advantage by avoiding the performance degradation caused by spatial-domain data biases in the two-stage approach. To illustrate the differences in diffusion tensors generated by different methods, the corresponding diffusion tensors are visualized in Fig. \ref{fig5}. As suggested in \cite{descoteaux2007regularized}, FA is used as the background for the visualization of diffusion tensors. The yellow box highlights the zoomed-in region. The red arrow points to the tensors showing differences. The color and size of the ellipsoids indicate that the results generated by our method are closer to the reference data, demonstrating that the diffusion tensors fitted by our method are more accurate. Overall, our method enhances the quality of diffusion images and improves the accuracy of downstream tasks.

% Section 5.2.2  FOD
\subsubsection{Comparison on FOD}\label{sec5.2.2}
\begin{figure}[htbp]
    \centering
    
\includegraphics[width=0.95\linewidth]{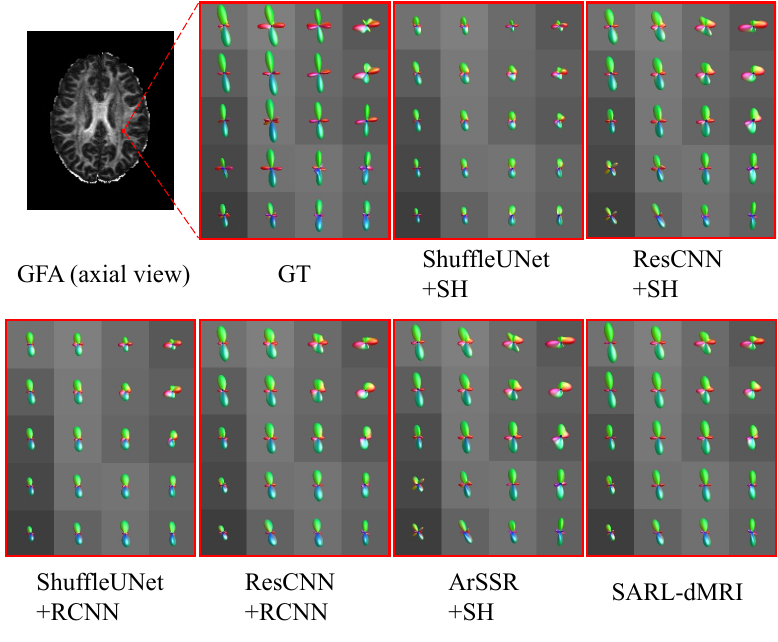}
    \caption{The qualitative comparison of FOD obtained by different methods. GFA measure is used as the background for FOD visualization. Red represents Left-Right (L-R), Green represents Anterior-Posterior (A-P), and Blue represents Superior-Inferior (S-I). The red box highlights the locally visualized area of the FOD.}\label{fig6}
\end{figure}

Diffusion images can be used to estimate the FOD in each voxel, allowing the observation of the general orientation of the fibers\cite{tournier2007robust}. Fig. \ref{fig6} shows the visualization results of the FOD obtained by different methods. Following the recommendation in \cite{tuch2004q}, we use generalized fractional anisotropy (GFA) as the background for FOD visualization. The red box highlights the FOD visualization in a locally magnified area. Compared to other methods, the FOD generated by our approach better aligns with the reference results and shows fewer error peaks. This indicates that our method can better preserve fiber orientation information.

% Section 5.3 Ablation Study
\subsection{Ablation Studies}\label{sec5.3}
We conduct ablation studies to evaluate the effectiveness of the proposed DFM and wavelet-based frequency loss $\ell_{d}$. Table \ref{tab4} presents the quantitative results of ablation studies under four different in-distribution scenarios. The experimental results demonstrate that the DFM and wavelet-based frequency loss $\ell_{d}$ effectively enhance the reconstruction quality of diffusion images compared to the baseline model, thereby validating their effectiveness. Moreover, the DFM minimizes the discrepancy between the generated diffusion images and the corresponding regions of the original input data, thus enhancing the overall quality of the generated images. The wavelet-based frequency loss enhances the recovery of image details by extracting high-frequency information across different scales. Therefore, combining the DFM and the wavelet-based frequency loss further improves the quality of the generated diffusion images.

%figure 6

% Section 6 Conclusion
\section{Conclusion}\label{sec6}
In this study, we propose a novel framework, SARL-dMRI, which simultaneously improves the spatial and angular resolution of dMRI data. Specifically, it explores implicit neural representations and spherical harmonics to learn continuous representations in both domains, enhancing its applicability. Furthermore, we incorporate a data fidelity module to guarantee that the generated diffusion images are consistent with the original input data. Meanwhile, we design a wavelet-based frequency loss to capture multi-scale high-frequency information, enhancing the recovery of image details. Extensive experiments demonstrate the effectiveness of SARL-dMRI, showcasing its ability to improve the resolution of diffusion images and enhance the performance of downstream tasks. Our method offers a new solution to accelerate the acquisition of high-resolution dMRI data, opening up new possibilities for future clinical applications.

\bibliographystyle{IEEEtran}
\bibliography{reference}

\begin{thebibliography}{10}
\providecommand{\url}[1]{#1}
\csname url@samestyle\endcsname
\providecommand{\newblock}{\relax}
\providecommand{\bibinfo}[2]{#2}
\providecommand{\BIBentrySTDinterwordspacing}{\spaceskip=0pt\relax}
\providecommand{\BIBentryALTinterwordstretchfactor}{4}
\providecommand{\BIBentryALTinterwordspacing}{\spaceskip=\fontdimen2\font plus
\BIBentryALTinterwordstretchfactor\fontdimen3\font minus \fontdimen4\font\relax}
\providecommand{\BIBforeignlanguage}[2]{{%
\expandafter\ifx\csname l@#1\endcsname\relax
\typeout{** WARNING: IEEEtran.bst: No hyphenation pattern has been}%
\typeout{** loaded for the language `#1'. Using the pattern for}%
\typeout{** the default language instead.}%
\else
\language=\csname l@#1\endcsname
\fi
#2}}
\providecommand{\BIBdecl}{\relax}
\BIBdecl

\bibitem{descoteaux2007regularized}
M.~Descoteaux, E.~Angelino, S.~Fitzgibbons, and R.~Deriche, ``\textrm{Regularized, fast, and robust analytical Q-ball imaging},'' \emph{Magnetic Resonance in Medicine}, vol.~58, no.~3, pp. 497--510, 2007.

\bibitem{tournier2011diffusion}
J.-D. Tournier, S.~Mori, and A.~Leemans, ``\textrm{Diffusion tensor imaging and beyond},'' \emph{Magnetic resonance in medicine}, vol.~65, no.~6, p. 1532, 2011.

\bibitem{martin2025conditional}
P.~Martin, M.~Altbach, and A.~Bilgin, ``\textrm{Conditional generative diffusion deep learning for accelerated diffusion tensor and kurtosis imaging},'' \emph{Magnetic Resonance Imaging}, vol. 117, p. 110309, 2025.

\bibitem{spagnolo2024down}
F.~Spagnolo, S.~Gobbi, E.~Zsoldos, M.~Edde, M.~Weigel, C.~Granziera, M.~Descoteaux, M.~Barakovic, and S.~Magon, ``\textrm{Down-sampling in diffusion MRI: a bundle-specific DTI and NODDI study},'' \emph{Frontiers in Neuroimaging}, vol.~3, p. 1359589, 2024.

\bibitem{cascio2007diffusion}
C.~J. Cascio, G.~Gerig, and J.~Piven, ``\textrm{Diffusion tensor imaging: application to the study of the developing brain},'' \emph{Journal of the American Academy of Child \& Adolescent Psychiatry}, vol.~46, no.~2, pp. 213--223, 2007.

\bibitem{sun2024disturbed}
S.~Sun, M.~Tian, X.~Lin, and P.~Zhao, ``\textrm{Disturbed white matter integrity on diffusion tensor imaging in young children with epilepsy},'' \emph{Clinical Radiology}, vol.~79, no.~1, pp. e119--e126, 2024.

\bibitem{eriksson2001diffusion}
S.~Eriksson, F.~Rugg-Gunn, M.~Symms, G.~Barker, and J.~Duncan, ``\textrm{Diffusion tensor imaging in patients with epilepsy and malformations of cortical development},'' \emph{Brain}, vol. 124, no.~3, pp. 617--626, 2001.

\bibitem{shen2024exploring}
Y.~Shen, X.~Zhao, K.~Wang, Y.~Sun, X.~Zhang, C.~Wang, Z.~Yang, Z.~Feng, and X.~Zhang, ``\textrm{Exploring White Matter Abnormalities in Young Children with Autism Spectrum Disorder: Integrating Multi-shell Diffusion Data and Machine Learning Analysis},'' \emph{Academic Radiology}, vol.~31, no.~5, pp. 2074--2084, 2024.

\bibitem{le2001diffusion}
D.~Le~Bihan, J.-F. Mangin, C.~Poupon, C.~A. Clark, S.~Pappata, N.~Molko, and H.~Chabriat, ``\textrm{Diffusion tensor imaging: concepts and applications},'' \emph{Journal of Magnetic Resonance Imaging: An Official Journal of the International Society for Magnetic Resonance in Medicine}, vol.~13, no.~4, pp. 534--546, 2001.

\bibitem{dubois2014early}
J.~Dubois, G.~Dehaene-Lambertz, S.~Kulikova, C.~Poupon, P.~S. H{\"u}ppi, and L.~Hertz-Pannier, ``\textrm{The early development of brain white matter: a review of imaging studies in fetuses, newborns and infants},'' \emph{neuroscience}, vol. 276, pp. 48--71, 2014.

\bibitem{yin2016unified}
S.~Yin, X.~You, W.~Xue, B.~Li, Y.~Zhao, X.-Y. Jing, P.~S. Wang, and Y.~Tang, ``\textrm{A unified approach for spatial and angular super-resolution of diffusion tensor MRI},'' in \emph{Pattern Recognition: 7th Chinese Conference, CCPR 2016, Chengdu, China, November 5-7, 2016, Proceedings, Part II 7}.\hskip 1em plus 0.5em minus 0.4em\relax Springer, 2016, pp. 312--324.

\bibitem{chen2019xq}
G.~Chen, B.~Dong, Y.~Zhang, W.~Lin, D.~Shen, and P.-T. Yap, ``\textrm{XQ-SR: joint xq space super-resolution with application to infant diffusion MRI},'' \emph{Medical image analysis}, vol.~57, pp. 44--55, 2019.

\bibitem{yin2018joint}
S.~Yin, X.~You, X.~Yang, Q.~Peng, Z.~Zhu, and X.-Y. Jing, ``\textrm{A joint space-angle regularization approach for single 4D diffusion image super-resolution},'' \emph{Magnetic Resonance in Medicine}, vol.~80, no.~5, pp. 2173--2187, 2018.

\bibitem{luo2022diffusion}
S.~Luo, J.~Zhou, Z.~Yang, H.~Wei, and Y.~Fu, ``\textrm{Diffusion MRI super-resolution reconstruction via sub-pixel convolution generative adversarial network},'' \emph{Magnetic Resonance Imaging}, vol.~88, pp. 101--107, 2022.

\bibitem{scherrer2011super}
B.~Scherrer, A.~Gholipour, and S.~K. Warfield, ``\textrm{Super-resolution in diffusion-weighted imaging},'' in \emph{Medical Image Computing and Computer-Assisted Intervention--MICCAI 2011: 14th International Conference, Toronto, Canada, September 18-22, 2011, Proceedings, Part II 14}.\hskip 1em plus 0.5em minus 0.4em\relax Springer, 2011, pp. 124--132.

\bibitem{wang2024knowledge}
S.~Wang, R.~Wu, S.~Jia, A.~Diakite, C.~Li, Q.~Liu, H.~Zheng, and L.~Ying, ``\textrm{Knowledge-driven deep learning for fast MR imaging: Undersampled MR image reconstruction from supervised to un-supervised learning},'' \emph{Magnetic Resonance in Medicine}, vol.~92, no.~2, pp. 496--518, 2024.

\bibitem{du2020super}
J.~Du, Z.~He, L.~Wang, A.~Gholipour, Z.~Zhou, D.~Chen, and Y.~Jia, ``\textrm{Super-resolution reconstruction of single anisotropic 3D MR images using residual convolutional neural network},'' \emph{Neurocomputing}, vol. 392, pp. 209--220, 2020.

\bibitem{alexander2017image}
D.~C. Alexander, D.~Zikic, A.~Ghosh, R.~Tanno, V.~Wottschel, J.~Zhang, E.~Kaden, T.~B. Dyrby, S.~N. Sotiropoulos, H.~Zhang \emph{et~al.}, ``\textrm{Image quality transfer and applications in diffusion MRI},'' \emph{NeuroImage}, vol. 152, pp. 283--298, 2017.

\bibitem{tanno2021uncertainty}
R.~Tanno, D.~E. Worrall, E.~Kaden, A.~Ghosh, F.~Grussu, A.~Bizzi, S.~N. Sotiropoulos, A.~Criminisi, and D.~C. Alexander, ``\textrm{Uncertainty modelling in deep learning for safer neuroimage enhancement: Demonstration in diffusion MRI},'' \emph{NeuroImage}, vol. 225, p. 117366, 2021.

\bibitem{chatterjee2021shuffleunet}
S.~Chatterjee, A.~Sciarra, M.~D{\"u}nnwald, R.~V. Mushunuri, R.~Podishetti, R.~N. Rao, G.~D. Gopinath, S.~Oeltze-Jafra, O.~Speck, and A.~N{\"u}rnberger, ``\textrm{ShuffleUNet: Super resolution of diffusion-weighted MRIs using deep learning},'' in \emph{2021 29th European Signal Processing Conference (EUSIPCO)}.\hskip 1em plus 0.5em minus 0.4em\relax IEEE, 2021, pp. 940--944.

\bibitem{wu2022arbitrary}
Q.~Wu, Y.~Li, Y.~Sun, Y.~Zhou, H.~Wei, J.~Yu, and Y.~Zhang, ``\textrm{An arbitrary scale super-resolution approach for 3d MR images via implicit neural representation},'' \emph{IEEE Journal of Biomedical and Health Informatics}, vol.~27, no.~2, pp. 1004--1015, 2022.

\bibitem{zhu2023super}
Z.~Zhu, J.~Ma, and C.~Zhang, ``\textrm{Super-resolution of diffusion MRI via joint consideration of spatial and angular spaces using graph CNN},'' \emph{Journal of Mechanics in Medicine and Biology}, vol.~23, no.~09, p. 2340095, 2023.

\bibitem{guo2023semi}
H.~Guo, L.~Wang, Y.~Gu, J.~Zhang, and Y.~Zhu, ``\textrm{Semi-supervised super-resolution of diffusion-weighted images based on multiple references},'' \emph{NMR in Biomedicine}, vol.~36, no.~8, p. e4919, 2023.

\bibitem{zhao2024super}
X.~Zhao and Z.~Wen, ``\textrm{Super-resolution of diffusion-weighted images using space-customized learning model},'' \emph{Technology and Health Care}, no. Preprint, pp. 1--13, 2024.

\bibitem{wu2024csr}
R.~Wu, J.~Cheng, C.~Li, J.~Zou, J.~Yang, W.~Fan, Y.~Liang, and S.~Wang, ``\textrm{CSR-dMRI: Continuous Super-Resolution of Diffusion MRI with Anatomical Structure-Assisted Implicit Neural Representation Learning},'' in \emph{International Workshop on Machine Learning in Medical Imaging}.\hskip 1em plus 0.5em minus 0.4em\relax Springer, 2024, pp. 114--123.

\bibitem{albay2018diffusion}
E.~Albay, U.~Demir, and G.~Unal, ``\textrm{Diffusion MRI spatial super-resolution using generative adversarial networks},'' in \emph{PRedictive Intelligence in MEdicine: First International Workshop, PRIME 2018, Held in Conjunction with MICCAI 2018, Granada, Spain, September 16, 2018, Proceedings 1}.\hskip 1em plus 0.5em minus 0.4em\relax Springer, 2018, pp. 155--163.

\bibitem{shi2016real}
W.~Shi, J.~Caballero, F.~Husz{\'a}r, J.~Totz, A.~P. Aitken, R.~Bishop, D.~Rueckert, and Z.~Wang, ``\textrm{Real-time single image and video super-resolution using an efficient sub-pixel convolutional neural network},'' in \emph{Proceedings of the IEEE conference on computer vision and pattern recognition}, 2016, pp. 1874--1883.

\bibitem{te2018rgcnn}
G.~Te, W.~Hu, A.~Zheng, and Z.~Guo, ``\textrm{RGCNN: Regularized graph cnn for point cloud segmentation},'' in \emph{Proceedings of the 26th ACM international conference on Multimedia}, 2018, pp. 746--754.

\bibitem{ma2021hybrid}
J.~Ma and H.~Cui, ``\textrm{Hybrid graph convolutional neural networks for super resolution of DW images},'' in \emph{Computational Diffusion MRI: International MICCAI Workshop, Lima, Peru, October 2020}.\hskip 1em plus 0.5em minus 0.4em\relax Springer, 2021, pp. 201--212.

\bibitem{wang2023uncertainty}
C.~Wang and J.~Ma, ``\textrm{Uncertainty-Supervised Super-Resolution Deep Learning Network in Diffusion MRI},'' \emph{Highlights in Science, Engineering and Technology}, vol.~45, pp. 7--10, 2023.

\bibitem{yin2019fast}
S.~Yin, Z.~Zhang, Q.~Peng, and X.~You, ``\textrm{Fast and accurate reconstruction of hardi using a 1d encoder-decoder convolutional network},'' \emph{arXiv preprint arXiv:1903.09272}, 2019.

\bibitem{lyon2022angular}
M.~Lyon, P.~Armitage, and M.~A. {\'A}lvarez, ``\textrm{Angular super-resolution in diffusion MRI with a 3D recurrent convolutional autoencoder},'' in \emph{International Conference on Medical Imaging with Deep Learning}.\hskip 1em plus 0.5em minus 0.4em\relax PMLR, 2022, pp. 834--846.

\bibitem{chen2023super}
Z.~Chen, C.~Peng, Y.~Li, Q.~Zeng, and Y.~Feng, ``\textrm{Super-resolved q-space learning of diffusion MRI},'' \emph{Medical Physics}, vol.~50, no.~12, pp. 7700--7713, 2023.

\bibitem{lyon2024spatio}
M.~Lyon, P.~Armitage, and M.~A. {\'A}lvarez, ``\textrm{Spatio-angular convolutions for super-resolution in diffusion MRI},'' \emph{Advances in Neural Information Processing Systems}, vol.~36, 2024.

\bibitem{weiss2021towards}
T.~Weiss, S.~Vedula, O.~Senouf, O.~Michailovich, and A.~Bronstein, ``\textrm{Towards learned optimal q-space sampling in diffusion MRI},'' in \emph{Computational Diffusion MRI: International MICCAI Workshop, Lima, Peru, October 2020}.\hskip 1em plus 0.5em minus 0.4em\relax Springer, 2021, pp. 13--28.

\bibitem{suzuki2024high}
Y.~Suzuki, T.~Ueyama, K.~Sakata, A.~Kasahara, H.~Iwanaga, K.~Yasaka, and O.~Abe, ``\textrm{High-angular resolution diffusion imaging generation using 3D U-Net},'' \emph{Neuroradiology}, vol.~66, no.~3, pp. 371--387, 2024.

\bibitem{liu2023accelerated}
S.~Liu, Y.~Liu, X.~Xu, R.~Chen, D.~Liang, Q.~Jin, H.~Liu, G.~Chen, and Y.~Zhu, ``\textrm{Accelerated cardiac diffusion tensor imaging using deep neural network},'' \emph{Physics in Medicine \& Biology}, vol.~68, no.~2, p. 025008, 2023.

\bibitem{wang2024ultrafast}
J.~Wang, Z.~Chen, C.~Cai, and S.~Cai, ``\textrm{Ultrafast diffusion tensor imaging based on deep learning and multi-slice information sharing},'' \emph{Physics in Medicine \& Biology}, vol.~69, no.~3, p. 035011, 2024.

\bibitem{karimi2022diffusion}
D.~Karimi and A.~Gholipour, ``\textrm{Diffusion tensor estimation with transformer neural networks},'' \emph{Artificial intelligence in medicine}, vol. 130, p. 102330, 2022.

\bibitem{tian2020deepdti}
Q.~Tian, B.~Bilgic, Q.~Fan, C.~Liao, C.~Ngamsombat, Y.~Hu, T.~Witzel, K.~Setsompop, J.~R. Polimeni, and S.~Y. Huang, ``\textrm{DeepDTI: High-fidelity six-direction diffusion tensor imaging using deep learning},'' \emph{NeuroImage}, vol. 219, p. 117017, 2020.

\bibitem{li2021superdti}
H.~Li, Z.~Liang, C.~Zhang, R.~Liu, J.~Li, W.~Zhang, D.~Liang, B.~Shen, X.~Zhang, Y.~Ge \emph{et~al.}, ``\textrm{SuperDTI: Ultrafast DTI and fiber tractography with deep learning},'' \emph{Magnetic resonance in medicine}, vol.~86, no.~6, pp. 3334--3347, 2021.

\bibitem{qin2021super}
Y.~Qin, Z.~Liu, C.~Liu, Y.~Li, X.~Zeng, and C.~Ye, ``\textrm{Super-Resolved q-Space deep learning with uncertainty quantification},'' \emph{Medical Image Analysis}, vol.~67, p. 101885, 2021.

\bibitem{chen2021learning}
Y.~Chen, S.~Liu, and X.~Wang, ``\textrm{Learning continuous image representation with local implicit image function},'' in \emph{Proceedings of the IEEE/CVF conference on computer vision and pattern recognition}, 2021, pp. 8628--8638.

\bibitem{molaei2023implicit}
A.~Molaei, A.~Aminimehr, A.~Tavakoli, A.~Kazerouni, B.~Azad, R.~Azad, and D.~Merhof, ``\textrm{Implicit neural representation in medical imaging: A comparative survey},'' in \emph{Proceedings of the IEEE/CVF International Conference on Computer Vision}, 2023, pp. 2381--2391.

\bibitem{shen2022nerp}
L.~Shen, J.~Pauly, and L.~Xing, ``\textrm{NeRP: implicit neural representation learning with prior embedding for sparsely sampled image reconstruction},'' \emph{IEEE Transactions on Neural Networks and Learning Systems}, vol.~35, no.~1, pp. 770--782, 2022.

\bibitem{zhang2021nerd}
H.~Zhang, R.~Wang, J.~Zhang, C.~Li, G.~Yang, P.~Spincemaille, T.~Nguyen, and Y.~Wang, ``\textrm{Nerd: Neural representation of distribution for medical image segmentation},'' \emph{arXiv preprint arXiv:2103.04020}, 2021.

\bibitem{anderson2005measurement}
A.~W. Anderson, ``\textrm{Measurement of fiber orientation distributions using high angular resolution diffusion imaging},'' \emph{Magnetic Resonance in Medicine: An Official Journal of the International Society for Magnetic Resonance in Medicine}, vol.~54, no.~5, pp. 1194--1206, 2005.

\bibitem{ha2022spharm}
S.~Ha and I.~Lyu, ``\textrm{SPHARM-Net: Spherical harmonics-based convolution for cortical parcellation},'' \emph{IEEE Transactions on Medical Imaging}, vol.~41, no.~10, pp. 2739--2751, 2022.

\bibitem{tournier2007robust}
J.-D. Tournier, F.~Calamante, and A.~Connelly, ``\textrm{Robust determination of the fibre orientation distribution in diffusion MRI: non-negativity constrained super-resolved spherical deconvolution},'' \emph{Neuroimage}, vol.~35, no.~4, pp. 1459--1472, 2007.

\bibitem{zhang2018residual}
Y.~Zhang, Y.~Tian, Y.~Kong, B.~Zhong, and Y.~Fu, ``\textrm{Residual dense network for image super-resolution},'' in \emph{Proceedings of the IEEE conference on computer vision and pattern recognition}, 2018, pp. 2472--2481.

\bibitem{zhao2016loss}
H.~Zhao, O.~Gallo, I.~Frosio, and J.~Kautz, ``\textrm{Loss functions for image restoration with neural networks},'' \emph{IEEE Transactions on computational imaging}, vol.~3, no.~1, pp. 47--57, 2016.

\bibitem{guo2017deep}
T.~Guo, H.~Seyed~Mousavi, T.~Huu~Vu, and V.~Monga, ``\textrm{Deep wavelet prediction for image super-resolution},'' in \emph{Proceedings of the IEEE conference on computer vision and pattern recognition workshops}, 2017, pp. 104--113.

\bibitem{glasser2013minimal}
M.~F. Glasser, S.~N. Sotiropoulos, J.~A. Wilson, T.~S. Coalson, B.~Fischl, J.~L. Andersson, J.~Xu, S.~Jbabdi, M.~Webster, J.~R. Polimeni \emph{et~al.}, ``\textrm{The minimal preprocessing pipelines for the Human Connectome Project},'' \emph{Neuroimage}, vol.~80, pp. 105--124, 2013.

\bibitem{tuch2004q}
D.~S. Tuch, ``\textrm{Q-ball imaging},'' \emph{Magnetic Resonance in Medicine: An Official Journal of the International Society for Magnetic Resonance in Medicine}, vol.~52, no.~6, pp. 1358--1372, 2004.

\end{thebibliography}

\end{document}